# Phase-field modeling of brittle fracture with multi-level $hp$-FEM and the finite cell method

S. Nagaraja[*,1], M. Elhaddad[2], M. Ambati[1],
S. Kollmannsberger[2], L. De Lorenzis[1], and E. Rank[2,3]

[1]*Institute of Applied Mechanics, Technische Universität Braunschweig, Pockelstr. 3, 38106 Braunschweig, Germany*
[2]*Chair for Computation in Engineering, Technische Universität München, Arcisstr. 21, 80333 München, Germany*
[3]*Institute for Advanced Study, Technische Universität München, Lichtenbergstr. 2a, 85748 Garching, Germany*

## Abstract

The difficulties in dealing with discontinuities related to a sharp crack are overcome in the phase-field approach for fracture by modeling the crack as a diffusive object being described by a continuous field having high gradients. The discrete crack limit case is approached for a small length-scale parameter that controls the width of the transition region between the fully broken and the undamaged phases. From a computational standpoint, this necessitates fine meshes, at least locally, in order to accurately resolve the phase-field profile. In the classical approach, phase-field models are computed on a fixed mesh that is a priori refined in the areas where the crack is expected to propagate. This on the other hand curbs the convenience of using phase-field models for unknown crack paths and its ability to handle complex crack propagation patterns. In this work, we overcome this issue by employing the multi-level $hp$-refinement technique that enables a dynamically changing mesh which in turn allows the refinement to remain local at singularities and high gradients without problems of hanging nodes. Yet, in case of complex geometries, mesh generation and in particular local refinement becomes non-trivial. We address this issue by integrating a two-dimensional phase-field framework for brittle fracture with the finite cell method (FCM). The FCM based on high-order finite elements is a non-geometry-conforming discretization technique wherein the physical domain is embedded into a larger fictitious domain of simple geometry that can be easily discretized. This facilitates mesh generation for complex geometries and supports local refinement. Numerical examples including a comparison to a validation experiment illustrate the applicability of the multi-level $hp$-refinement and the FCM in the context of phase-field simulations.

## Contents

# 1 Introduction

The ability to accurately predict fracture in structures for industrial applications involving complex geometries and loading conditions is constantly gaining importance. Due to the nonlinear nature of fracture processes, in general it becomes inevitable to use numerical methods to solve fracture problems. In computational fracture mechanics, the virtual crack closure technique [43, 68, 71], the extended finite element method [55], cohesive zone methods [84, 81] and other discrete fracture models represent a crack as a discontinuous entity by virtue of which a discontinuity in the primary field variable (displacement field) is obtained. These models not only require re-meshing strategies or enrichment of the displacement field to accurately capture the fracture process, but also a robust algorithm to track the crack during the course of its propagation. This is numerically expensive and poses implementational challenges, especially in cases with complex crack patterns.

The phase-field approach to fracture which unfolded from the pioneering work of Francfort and Marigo [30] followed by its regularization by Bourdin et al. [16], (see [4] for a comprehensive review) leads to an elegant numerical technique wherein the implementational complexities due to discrete cracks are eliminated [17, 8, 46]. Here, a discrete crack is approximated by the steep local variation of a continuous scalar field called the *crack phase-field* or simply the *phase-field*. The evolution of the phase-field, governed by a variational principle, models the process of fracture. Thus, phenomena like crack propagation, kinking, branching and merging involving complex crack topologies are straightforward to simulate using phase-field models. This attractive feature has inspired various studies treating e.g. dynamic brittle fracture [18, 39, 70, 15], mixed-mode fracture [82], ductile fracture [3, 48, 5, 51, 6, 14], cohesive fracture [49, 79], fracture in anisotropic materials [47, 22, 74], fracture of plates and shells [7, 42], fracture in thin films [50], fracture coupled with diffusion [83], fracture in porous media [20, 35, 54], fracture phenomena in soft biological tissues [62, 34, 33], to mention a few.

In phase-field models, a discrete crack is transformed into a thin region across which the transition happens between damaged and undamaged states, and the width of this region is controlled by a length-scale parameter. Diminishing values of this parameter lead to the convergence of a phase-field crack to a sharp crack topology, but induces locally high gradients. This calls for finite element meshes with high spatial resolution. A uniform global $h$-refinement until the length-scale parameter is well resolved would be the most trivial option. However, this leads to excessive computational cost, calling for adaptive $h$-refinement [77]. To sidestep this to a certain extent, often employed is a fixed mesh with local refinements in the regions where the crack is expected to develop, as e.g. in [4, 15]. However, only in limited cases is the crack path a priori known. For these reasons, it becomes inevitable to opt for an adaptive refinement method for which no pre-knowledge of the crack pattern is necessary. In Borden et al. [15], an adaptive scheme has been developed and applied to dynamic fracture using T-spline-based isogeometric analysis. This refinement is, however, non-dynamic in nature, in the sense that the mesh does not change over the course of the simulation. The refinement is done once a certain ter-mination point is reached. This limits the range of its applicability and calls for a dynamic update of the local refinement. Chakraborty et al. [21] present a quasi-static example with dynamic $h$-adaptivity. In a very recent work, Patil et al [61] present local refinements in the context of adaptive multiscale phase field method. In Badnava et al. [10], a predictor-corrector scheme for mesh adaptivity proposed in [36] has been used for dynamic $h$-refinement. Yet these approaches are limited to linear finite elements. Extensions to high-order dynamic refinements have only appeared very recently [38, 37] and were published in the context of isogeometric analysis.

Furthermore, structures with complex geometries, which are common in engineering practice, require the definition of a boundary conforming finite element mesh. Generation of such meshes is one of the most time consuming and, in particular in 3D, most error-prone steps in a finite element analysis. Moreover, unstructured meshes are more difficult to dynamically re-mesh during the course of a simulation. This complicates the desired local refinement for problems involving non-trivial geometries.

In this work, we address both these challenges by employing the multi-level $hp$-refinement scheme: local refinement which is easy to dynamically update in conjunction with the finite cell method (FCM) to avoid meshing. The FCM introduced in [59, 25] is an embedded domain method. This non-boundary conforming discretization combines the advantages of immersed boundary methods and high-order finite elements. Here, the representation of the physical domain is separated from the approximation of the solution. A trivial mesh can be chosen, thus avoiding a numerically expensive and time consuming meshing procedure. In the simplest case this is an axis-aligned grid of squares or cubes. The original geometry is then observed during integration of the weak form, thereby recovering the favorable convergence properties of the underlying high-order discretization. This idea enables complex geometry representations and is also convenient when the structure under consideration or loading are changing with time [85]. In recent works, the FCM has been combined with the multi-level $hp$-refinement scheme in Zander et al. [89], Zander [85], Bog et al. [13], Elhaddad et al. [27] to support adaptive mesh refinement.

Furthermore, in parts of the domain where the solution is smooth, i.e. distant from the crack or geometry and load induced singularities, the use of high-order shape functions is an appealing alternative. The felicitous multi-level $hp$-refinement scheme exploits the advantages of both the $h$- and $p$-extension of the finite element method (FEM).

This paper is structured as follows: The investigation starts in section 2 with a brief overview of the underlying phase-field model for brittle fracture. Section 3 firstly presents the numerical formulation of the coupled partial differential equations (PDEs) governing the fracture problem solved using the FEM and a staggered solution scheme [31]. This is followed by a brief section about the multi-level $hp$-refinement which features the employed refinement strategy. Section 3 is then concerned with the basic concept of the FCM. In section 4, several numerical examples including quasi-static and dynamic cases are considered. It is shown that the large gradients in the phase-field can be used to drive a dynamic refinement (a mesh that changes over the course of the simulation) without any additional criteria for crack propagation and branching (in dynamic fracture). The final numerical example showcases an application of the finite cell discretization technique for phase-field models including experimental validation of the results. A most important observation is that the propagating cracks can readily enter into or start from voids in a structure, without the necessity to mesh the boundary of the voids by conforming finite elements. This opens the door to efficiently simulating fracture in highly complex domains like porous structures. Finally, we provide conclusions of the article in section 5.

## 2 Phase-field modeling of brittle fracture

We devote this section to the phase-field modeling of dynamic brittle fracture. Here, we adopt the rate-independent dynamic phase-field model introduced by Borden et al. [15] which essentially is an extension of the quasi-static model

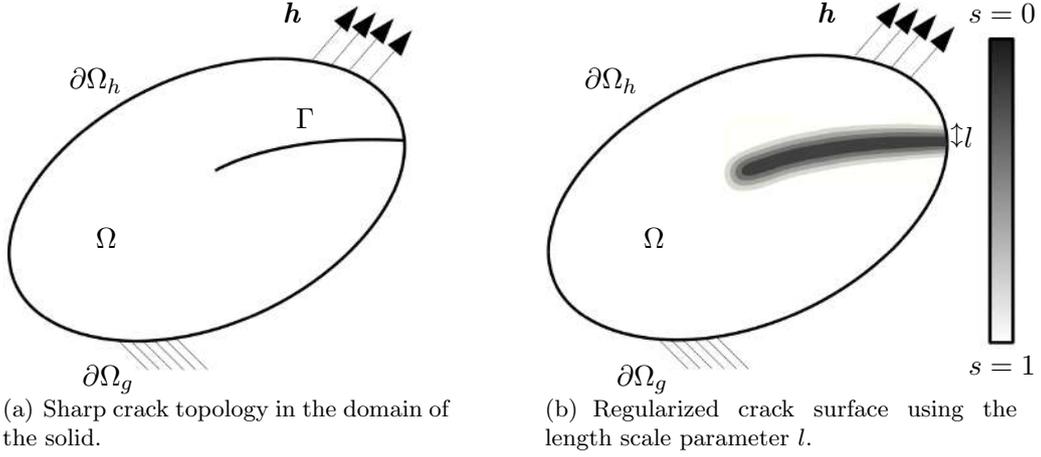

(a) Sharp crack topology in the domain of the solid.

(b) Regularized crack surface using the length scale parameter $l$.

Figure 1: Phase-field approximation of crack topology.

presented by Bourdin et al. [16].

Consider an arbitrarily shaped domain $\Omega \subset \mathbb{R}^d$, with $d$ as the spatial dimension, having boundary $\partial\Omega$. Suppose that this body contains a crack set denoted by $\Gamma$ as shown in Fig.1(a). Let $\boldsymbol{u}(\boldsymbol{x},t) \in \mathbb{R}^d$ be the displacement of a point $\boldsymbol{x} \in \Omega$ at time $t$. The displacement field satisfies the Dirichlet and Neumann boundary conditions on $\partial\Omega_g$ and $\partial\Omega_h$, respectively, with $\partial\Omega_g \cup \partial\Omega_h = \partial\Omega$ and $\partial\Omega_h \cap \partial\Omega_g = \emptyset$.

The total potential energy of the body is the sum of the elastic energy and the fracture energy,

$$\Psi_{pot}(\boldsymbol{u},\Gamma) = \int_\Omega \Psi_e(\boldsymbol{\varepsilon}(\boldsymbol{u}))d\boldsymbol{x} + \int_\Gamma G_c d\boldsymbol{s}. \quad (1)$$

Here, $\boldsymbol{\varepsilon}(\boldsymbol{x},t)$ is the standard second-order infinitesimal strain tensor defined as the symmetric part of the displacement gradient, $\Psi_e$ is the elastic strain energy density function and $G_c$ is the fracture toughness of the material. Similarly, the kinetic energy of the body is defined as

$$\Psi_{kin}(\dot{\boldsymbol{u}}) = \frac{1}{2}\int_\Omega \rho \dot{\boldsymbol{u}} \cdot \dot{\boldsymbol{u}} d\boldsymbol{x}, \quad (2)$$

with $\dot{\boldsymbol{u}} = \frac{\partial \boldsymbol{u}}{\partial t}$ as the velocity and $\rho$ as the mass density of the material. The Lagrangian for this problem is defined as

$$L(\boldsymbol{u},\dot{\boldsymbol{u}},\Gamma) = \Psi_{kin}(\dot{\boldsymbol{u}}) - \Psi_{pot}(\boldsymbol{u},\Gamma) \quad (3)$$

$$= \int_\Omega \left[\frac{1}{2}\rho\dot{\boldsymbol{u}}\cdot\dot{\boldsymbol{u}} - \Psi_e(\boldsymbol{\varepsilon}(\boldsymbol{u}))\right]d\boldsymbol{x} - \int_\Gamma G_c d\boldsymbol{s}. (4)$$

The variational formulation for brittle fracture introduced by Francfort and Marigo [30] is regularized using the phase-field approximation by Bourdin et al. [16] to enable a numerical implementation. In the regularized formulation, the fracture energy is approximated as

$$\int_\Gamma G_c d\boldsymbol{s} \approx \int_\Omega G_c \left[\frac{1}{4l}(1-s)^2 + l|\nabla_{,\boldsymbol{x}} s|^2\right]d\boldsymbol{x}. \quad (5)$$

The phase-field $s(\boldsymbol{x}, t)$ regularizes the representation of the crack and has a value of 0 in a fully cracked region and 1 at intact material points as shown in Fig.1(b). The length-scale parameter $l$ controls the width of the diffusive crack. Phase-field models converge to the well known Griffith's theory for brittle fracture (in the sense of $\Gamma$-convergence) as this parameter tends to zero [17].

To realize the coupling between the displacement field $\boldsymbol{u}$ and the phase-field $s$, the elastic strain energy density in the regularized formulation is given by

$$\Psi_e(\boldsymbol{\varepsilon}, s) = g(s)\Psi_e^+(\boldsymbol{\varepsilon}) + \Psi_e^-(\boldsymbol{\varepsilon}). \quad (6)$$

Here, $g(s)$ is called the *stress degradation function*. In order to achieve $\Gamma$-convergence as explained by Braides [19] and to ensure reasonable evolution of the stress and the crack fields, an appropriate stress degradation function has to be chosen. From the literature, $g(s) = (1-\eta)s^2 + \eta$, which satisfies these requirements, is used in this work [15]. Here, $\eta$ is a dimensionless parameter used to model a very small artificial stiffness of a completely damaged phase, typically of the order $10^{-6}$ [53]. $\Psi_e^+$ and $\Psi_e^-$ are the tensile and compressive components of the elastic strain energy density. In order to inhibit crack growth under compression and to distinguish between the fracture behavior in tension and compression, a selective degradation of the strain energy density, wherein only the tensile part is degraded by multiplication with the stress degradation function, is performed. Selective degradation of the strain energy density is important for the dynamic case in particular, since the stress waves reflecting from domain boundaries may otherwise create unphysical fracture patterns [15]. In this work, we adopt the tension-compression split proposed by Miehe et al. [52], where:

$$\Psi_e^+ = \frac{1}{2}\lambda \langle tr(\boldsymbol{\varepsilon}) \rangle^2 + \mu tr[(\boldsymbol{\varepsilon}^+)^2] \quad (7)$$

$$\Psi_e^- = \frac{1}{2}\lambda \big[(tr(\boldsymbol{\varepsilon}) - \langle tr(\boldsymbol{\varepsilon}) \rangle)^2\big] + \mu tr[(\boldsymbol{\varepsilon} - \boldsymbol{\varepsilon}^+)^2]. \quad (8)$$

Here, $\boldsymbol{\varepsilon}^+$ and $\boldsymbol{\varepsilon}^-$ are the tensile and compressive components of the strain tensor obtained by the spectral decomposition of $\boldsymbol{\varepsilon}$, whereas $\lambda$ and $\mu$ are the Lamé constants and

$$\langle X \rangle = \begin{cases} X & : X > 0 \\ 0 & : X \leq 0. \end{cases} \quad (9)$$

Following the work of Borden et al. [15], the Lagrangian for the regularized problem is thus defined as

$$L_l(\boldsymbol{u}, \dot{\boldsymbol{u}}, s) = \int_\Omega \left[ \frac{1}{2}\rho \dot{\boldsymbol{u}} \cdot \dot{\boldsymbol{u}} - g(s)\Psi_e^+ - \Psi_e^- \right] d\boldsymbol{x} \\ - \int_\Omega G_c \left[ \frac{1}{4l}(1-s)^2 + l|\nabla_{,\boldsymbol{x}}s|^2 \right] d\boldsymbol{x}. \quad (10)$$

Minimization of the Lagrangian in Eq:(10) leads to the governing equations in the strong form:

$$\begin{cases} \text{div}\boldsymbol{\sigma} + \rho \boldsymbol{b} = \rho \ddot{\boldsymbol{u}} & \text{on } \Omega \times ]0, T[ \\ \left[\frac{4l(1-\eta)}{G_c}\mathcal{H} + 1\right]s - 4l^2 \Delta s = 1 & \text{on } \Omega \times ]0, T[ \end{cases} \quad (11)$$

where $\boldsymbol{b}$ is the body force vector and the stress tensor $\boldsymbol{\sigma}$ is defined as

$$\boldsymbol{\sigma} = g(s)\frac{\partial \Psi_e^+}{\partial \boldsymbol{\varepsilon}} + \frac{\partial \Psi_e^-}{\partial \boldsymbol{\varepsilon}} \quad (12)$$

and

$$\mathcal{H} = \max_{\tau \in [0,t]} \Psi_e^+(\boldsymbol{\varepsilon}, \tau) \quad (13)$$

is a history variable introduced to enforce irreversibility of the phase-field evolution within a staggered solution scheme [52, 53].

The governing equations are complemented with the following boundary and initial conditions

$$\begin{cases} \boldsymbol{u}(\boldsymbol{x}, t) = \boldsymbol{g}(\boldsymbol{x}, t) & \text{on } \partial\Omega_g \times ]0, T[ \\ \boldsymbol{\sigma}(\boldsymbol{x}, t) \cdot \boldsymbol{n}(\boldsymbol{x}, t) = \boldsymbol{h}(\boldsymbol{x}, t) & \text{on } \partial\Omega_h \times ]0, T[ \\ \frac{\partial s}{\partial \boldsymbol{x}} \cdot \boldsymbol{n} = \boldsymbol{0} & \text{on } \partial\Omega \times ]0, T[ \end{cases} \quad (14)$$

$$\begin{cases} \boldsymbol{u}(\boldsymbol{x}, 0) = \boldsymbol{u}_0(\boldsymbol{x}) & \forall \boldsymbol{x} \in \Omega \\ \dot{\boldsymbol{u}}(\boldsymbol{x}, 0) = \dot{\boldsymbol{u}}_0(\boldsymbol{x}) & \forall \boldsymbol{x} \in \Omega. \end{cases} \quad (15)$$

Note that the formulation for quasi-static brittle fracture is a special case of the above if all the time dependent terms are omitted.

## 3 Numerical formulation

This section is dedicated to the weak formulation of the governing equations Eq:(11) first in a continuous form and then in the Galerkin setting in order to facilitate the finite element implementation. Subsequently, the basic concepts of high-order FEM and the multi-level $hp$-scheme used for adaptive refinement are summarized. Finally, the FCM is briefly introduced along with the finite cell formulation of the phase-field brittle fracture model.

### 3.1 Weak form and FEM

#### 3.1.1 Weak form

First, necessary function spaces for the derivation of the weak form are defined. Let $\boldsymbol{S}_t$ be the trial space for the displacement solution and $\tilde{S}_t$ the trial space for the phase-field solution. Similarly let $\boldsymbol{\mathcal{V}}$ and $\tilde{\mathcal{V}}$ be the corresponding test function spaces, as follows

$$\boldsymbol{S}_t = \{\boldsymbol{u}(\boldsymbol{x}, t) \in [H^1(\Omega)]^d | \boldsymbol{u} = \boldsymbol{g} \text{ on } \partial\Omega_g\} \quad (16)$$

$$\tilde{S}_t = \{s(\boldsymbol{x}, t) \in H^1(\Omega)\} \quad (17)$$

$$\boldsymbol{\mathcal{V}} = \{\boldsymbol{w}(\boldsymbol{x}) \in [H^1(\Omega)]^d | \boldsymbol{w} = \boldsymbol{0} \text{ on } \partial\Omega_g\} \quad (18)$$

$$\tilde{\mathcal{V}} = \{q(\boldsymbol{x}) \in H^1(\Omega)\} \quad (19)$$

$\forall t \in [0, T]$, where $H^1$ is the Sobolev space of first order [32]. The weak form is obtained by multiplying the strong form by admissible test functions and then applying Green's theorem. It reads as follows: given $\boldsymbol{g}$, $\boldsymbol{h}$, $\boldsymbol{u}_0$, $\dot{\boldsymbol{u}}_0$ and $s_0$, find $\boldsymbol{u}(\boldsymbol{x}, t) \in \boldsymbol{S}_t$ and $s(\boldsymbol{x}, t) \in \tilde{S}_t$ such that $\forall \mathbf{w} \in \boldsymbol{\mathcal{V}}, \forall q \in \tilde{\mathcal{V}}$

$$
\begin{aligned}
&(\rho \ddot{\boldsymbol{u}}, \boldsymbol{w})_\Omega + (\boldsymbol{\sigma}, \nabla \boldsymbol{w})_\Omega = (\rho \boldsymbol{b}, \boldsymbol{w})_\Omega + (\boldsymbol{h}, \boldsymbol{w})_{\partial \Omega_h} \\
&\left(\left[\tfrac{4l}{G_c}(1-\eta)\mathcal{H} + 1\right]s, q\right)_\Omega + (4l^2 \nabla s, \nabla q)_\Omega = (1, q)_\Omega \\
&(\rho \boldsymbol{u}(0), \boldsymbol{w})_\Omega = (\rho \boldsymbol{u}_0, \boldsymbol{w})_\Omega \\
&(\rho \dot{\boldsymbol{u}}(0), \boldsymbol{w})_\Omega = (\rho \dot{\boldsymbol{u}}_0, \boldsymbol{w})_\Omega \\
&(s(0), q)_\Omega = (s_0, q)_\Omega
\end{aligned}
\quad (20)
$$

where $(.,.)_\Omega$ is the $L_2$ inner product on $\Omega$ [32].

### 3.1.2 Galerkin form

Let $\boldsymbol{S}_t^h \subset \boldsymbol{S}_t$, $\tilde{S}_t^h \subset \tilde{S}_t$, $\boldsymbol{\mathcal{V}}^h \subset \boldsymbol{\mathcal{V}}$ and $\tilde{\mathcal{V}}^h \subset \tilde{\mathcal{V}}$ be finite dimensional approximating spaces. The semi-discrete (only spatially discretized) Galerkin form reads: find $\boldsymbol{u}^h(\boldsymbol{x}, t) \in \boldsymbol{S}_t^h$ and $s^h(\boldsymbol{x}, t) \in \tilde{S}_t^h$ such that

$$
\begin{aligned}
&(\rho \ddot{\boldsymbol{u}}^h, \boldsymbol{w}^h)_{\Omega^h} + (\boldsymbol{\sigma}, \nabla \boldsymbol{w}^h)_{\Omega^h} = (\rho \boldsymbol{b}^h, \boldsymbol{w}^h)_{\Omega^h} + (\boldsymbol{h}^h, \boldsymbol{w}^h)_{\partial \Omega_h^h} \\
&\left(\left[\tfrac{4l}{G_c}(1-\eta)\mathcal{H} + 1\right]s^h, q^h\right)_{\Omega^h} + (4l^2 \nabla s^h, \nabla q^h)_{\Omega^h} = (1, q^h)_{\Omega^h} \\
&(\rho \boldsymbol{u}^h(0), \boldsymbol{w}^h)_{\Omega^h} = (\rho \boldsymbol{u}_0, \boldsymbol{w}^h)_{\Omega^h} \\
&(\rho \dot{\boldsymbol{u}}^h(0), \boldsymbol{w}^h)_{\Omega^h} = (\rho \dot{\boldsymbol{u}}_0, \boldsymbol{w}^h)_{\Omega^h} \\
&(s^h(0), q^h)_{\Omega^h} = (s_0, q^h)_{\Omega^h}.
\end{aligned}
\quad (21)
$$

In the FEM, the domain $\Omega$ is approximated by $\Omega^h$ which consists of a finite number of elements. The discrete solutions, displacement $\boldsymbol{u}^h(\boldsymbol{x}, t)$ and phase field $s^h(\boldsymbol{x}, t)$, and the trial functions, $\boldsymbol{w}^h(\boldsymbol{x})$ for the displacement field and $q^h(\boldsymbol{x})$ for the phase field, are expressed as linear combinations of the basis functions. Here, the Bubnov-Galerkin formulation is followed, where the same basis functions are used for both the solution and the trial functions, hence

$$\boldsymbol{u}(\boldsymbol{x}, t) \approx \boldsymbol{u}^h(\boldsymbol{x}, t) = \sum_{A}^{n_b} N_A(\boldsymbol{x}) \boldsymbol{d}_A(t) \quad (22)$$

$$\boldsymbol{w}(\boldsymbol{x}) \approx \boldsymbol{w}^h(\boldsymbol{x}) = \sum_{A}^{n_b} N_A(\boldsymbol{x}) \boldsymbol{c}_A \quad (23)$$

$$s(\boldsymbol{x}, t) \approx s^h(\boldsymbol{x}, t) = \sum_{A}^{n_b} N_A(\boldsymbol{x}) \phi_A(t) \quad (24)$$

$$q(\boldsymbol{x}) \approx q^h(\boldsymbol{x}) = \sum_{A}^{n_b} N_A(\boldsymbol{x}) \chi_A, \quad (25)$$

where $n_b$ is the dimension of the discrete spaces, $N_A$ are the basis functions, $\boldsymbol{d}_A$, $\boldsymbol{c}_A$, $\phi_A$, $\chi_A$ are their coefficients. The semi-discrete form is to be discretized in time. A staggered time integration scheme is followed where the momentum and the phase-field equations are solved independently. The same predictor-corrector time integration method as in [15] is applied in this work.

### 3.2 Multi-level $hp$-FEM

In this section, the basic concept of the multi-level $hp$-FEM is explained. This refinement technique combines the advantages of the $h$- and $p$-version allowing high convergence rates even in the presence of high gradients or singularities in the solution field [85], which is the case in the proximity of the crack in fracture problems. While classical FEM is generally implemented using Lagrange shape functions which by nature are nodal basis functions, the $p$-version [9] makes use of integrated Legendre polynomials which are hierarchical in nature. This implies that there is an addition of one shape function every time the order of the polynomial is increased by one, while all other shape functions remain the same. Furthermore, the degrees of freedom may be directly assigned to the topological entities of the mesh: nodes, edges, faces and the interior of elements.

In classical $hp$-FEM, elements in the vicinity of high gradients are *replaced* by smaller elements of lower polynomial order. Elements away from the sensitive zone remain large and high-order polynomials are used to approximate the smooth part of the solution. This improves the solution accuracy significantly, compared to uniform $h$- or $p$-refinement with a comparable number of unknowns. Despite its excellent convergence behaviour, $hp$-FEM is not a popular method in engineering applications. This might at least partially be attributed to the fact that its implementation can be very complex [72], especially for mesh topologies allowing $n$-level hanging nodes. These hanging nodes are to be constrained appropriately to yield at least $C^0$-continuity in the numerical solution.

A remedy to this issue is offered by the multi-level $hp$-scheme introduced in [87]. The idea here is to *superpose* a base mesh with a finer overlay mesh at those locations of the domain where there are singularities or high gradients in the solution. This method mitigates the implementational complexity related to hanging nodes while constructing similar finite element spaces and thus retaining the same approximation quality as that from conventional $hp$-FEM [76]. The idea of *refinement-by-superposition* dates back to the pioneering work of Mote [57] and has been applied also in, for e.g. [63, 69, 11, 12, 29, 56]. While fine linear $h$ elements are superposed on a coarse $p$-FEM mesh in [63] and a hierarchic $hp$-$d$ scheme in which the base mesh is superposed by several layers of linear meshes is implemented in [69], the multi-level $hp$-scheme extends the idea of hierarchic overlay by introducing *high-order overlay meshes*. In this scheme, fine overlay meshes carry the high-order basis instead of the base mesh as shown in Fig.2. This feature of the multi-level $hp$-scheme leads to high convergence rates for problems with singularities and high gradients.

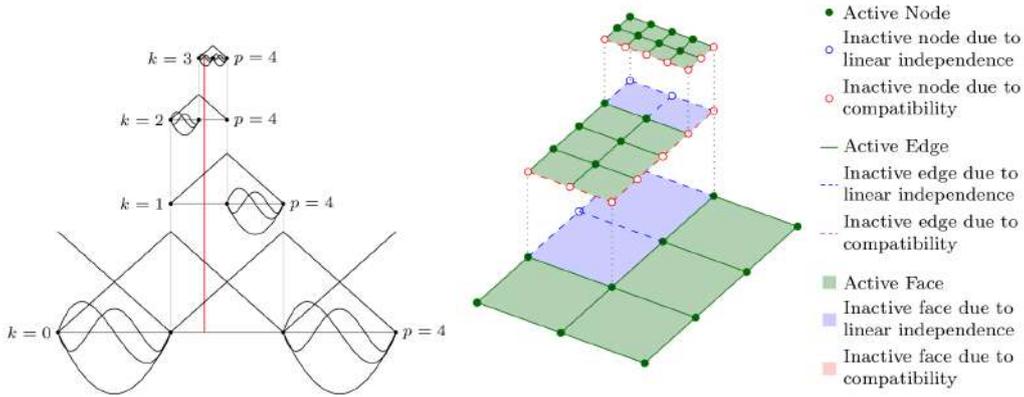

Figure 2: The multi-level $hp$-FEM concept in 1D and 2D following Zander et al. [87]

Another important feature of the multi-level $hp$-method is its archetype with a flexible data structure that allows for dynamic update of the discretization during the simulation as shown by [89, 13, 85]. This vital ingredient of the multi-level $hp$-method makes it applicable for the subject of this work. Furthermore, its advantage over classical $hp$-refinements lies in the fact that a simple set of rules exists which allows for a systematic, selective deactivation of those topological entities which either lead to linear dependencies or cause a violation of the globally required $C^0$-continuity. To this end, compatibility is achieved by imposing homogeneous Dirichlet boundary conditions on each layer of superposed overlay mesh, thus maintaining $C^0$ continuity. This corresponds to deactivating those components that are connected to the boundary of the overlay mesh. In 1D nodes, in 2D nodes and edges, in 3D nodes, edges and faces on the overlay boundary are to be deactivated. Linear independence of the basis is achieved by deactivating all topological components that have active sub-components. That is, high-order shape functions are deactivated on those elements which have an underlying child element. This concept is elucidated in Fig.2, where $k$ refers to the refinement depth and $p$ is the ansatz order. This also enables an optimal refinement using finer elements (with lower polynomial orders in the non-uniform refinement version) in the adjacency of singularities and strong gradients to ensure small discretization errors. All details can be found in [87].

### 3.2.1 Refinement strategy for fracture problems

To achieve locally bound refinement zones in order to ensure a better quality of the solution for a given computational cost (or to reduce the computational cost for a given quality), we integrate the phase-field model with the multi-level $hp$-FEM refinement scheme. Adopted here is the uniform version of the multi-level $hp$-FEM in which $h$-refinement is performed in the regions around the crack where high gradients in the solution exist, while the polynomial order remains constant at all refinement levels. Here,

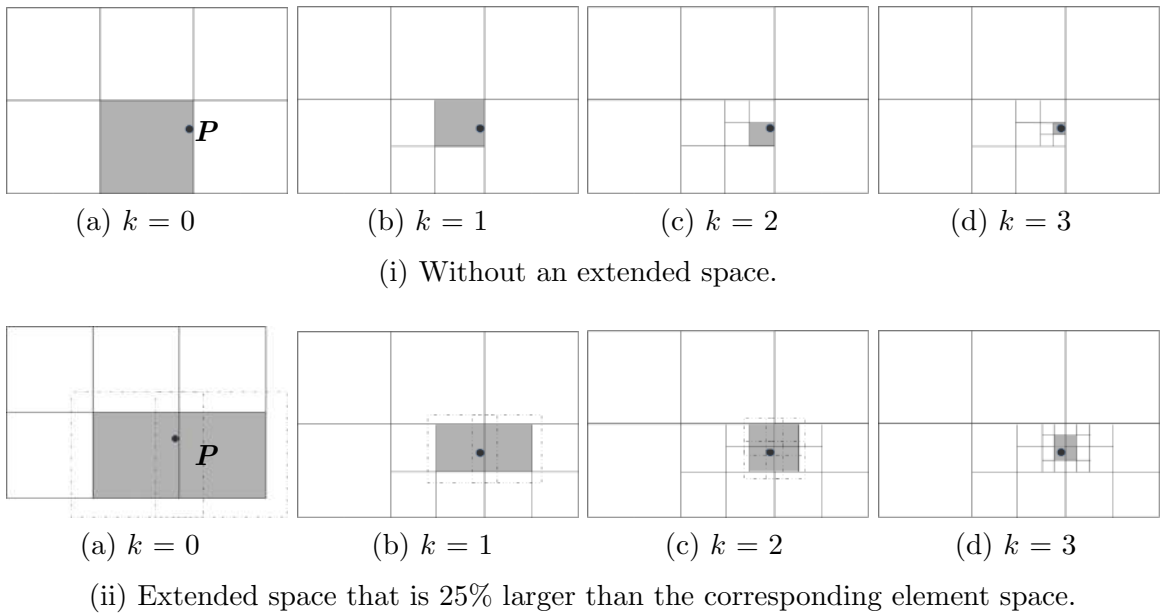

(a) $k = 0$  (b) $k = 1$  (c) $k = 2$  (d) $k = 3$

(i) Without an extended space.

(a) $k = 0$  (b) $k = 1$  (c) $k = 2$  (d) $k = 3$

(ii) Extended space that is 25% larger than the corresponding element space.

Figure 3: Geometric grading of mesh towards a singularity represented by $\boldsymbol{P}$.

two separate yet identical discretizations are defined, one for the elastic problem (momentum equation) and one for the phase-field problem (phase-field evolution equation). In every load (or time) step, first the elastic mesh is refined based on the phase-field solution. The phase-field parameter is a good indicator as to where refinements are to be carried out, since it undergoes steep variations in the vicinity of the crack and has a value close to one everywhere away from the crack. This avoids the use of algorithmically and mathematically much more engaged error estimators at practically no additional cost. A threshold value can be chosen for the phase-field parameter below which a refinement has to be carried out. Experience shows that a threshold between 0.75 and 0.85 governs the farthest elements from the crack to remain large, ensuring efficient refinement. Refinements are carried out until a predefined maximum is reached. The phase-field mesh then simply follows the elastic mesh refinement.

One of the fundamental principles in a *hp*-scheme along with order elevation of the basis is to geometrically grade the mesh towards the singularities and high gradients: decrease the element size close to the singularities and high gradients [85]. This can be robustly achieved by considering an extended space around the element for refinement, and sampling this extended space for values that overshoot the specified threshold criterion. The element and its neighbouring elements that fall in the range decided by the size of the extended space are recursively refined to obtain a graded mesh as shown in Fig.3(ii)(a)-(d). In Fig.3, the grey area indicates the marked elements for further refinement. The importance of using an extended space is highlighted in Fig.3(i)(d) where the element in the immediate vicinity of the singularity remains large which could lead to undesired numerical effects.

### 3.3 Finite cell method

The FCM is an immersed boundary method that simplifies the mesh generation process, maintaining the optimal convergence rate for smooth problems. For piecewise-smooth problems, a weak coupling approach recovers the optimal convergence [66]. For singular problems, a combination with multi-level *hp*-refinement yields a significant improvement in the convergence rates [27]. The FCM has been applied in solid mechanics in the context of e.g. elastoplasticity [2], elastodynamics [28, 40, 24], non-linear continuum mechanics [69], cohesive fracture [89], and thin-walled shell-like structures [64]. It has also been tested in various other fields, for example, for biomechanical applications [67, 80], numerical homogenization [26], topology optimization [60] multi-physics [88]. A MATLAB toolbox presented in [86] offers an easy entry point to the topic. In this work, the phase-field model for brittle fracture outlined in section 2 is combined with the FCM.

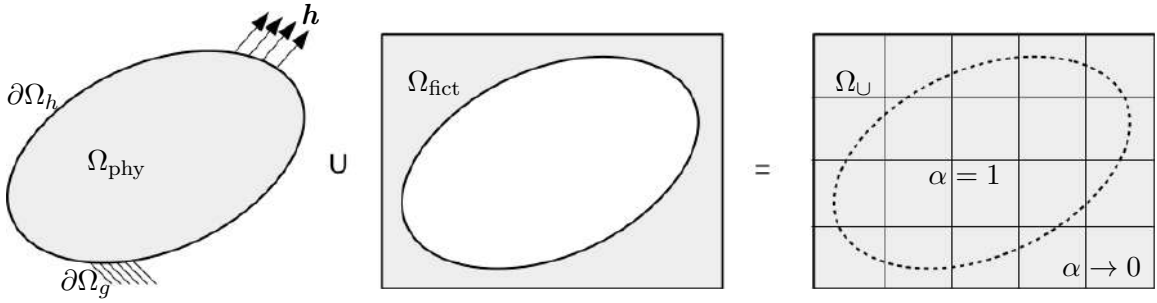

Figure 4: Basic concept of the finite cell method following [59].

### 3.3.1 Basic concept

To recover the original boundary value problem on $\Omega_\cup$, an indicator function $\alpha(\boldsymbol{x})$ is defined as

$$\alpha(\boldsymbol{x}) = \begin{cases} 1 & : \forall \boldsymbol{x} \in \Omega_{\text{phy}} \\ \epsilon \ll 1 & : \forall \boldsymbol{x} \notin \Omega_{\text{phy}}, \end{cases} \quad (26)$$

with $\epsilon$ being a small value but large enough to avoid ill-conditioning of the stiffness matrices of the elements that belong to the fictitious domain. This value introduces a controllable modeling error (in the order of $\sqrt{\epsilon}$ in the energy norm ) that, however, ensures numerical stability [23]. The weak form is multiplied by this indicator function to limit the contributions from the fictitious domain. This shifts the difficulty of resolving the geometry from the discretization of the primary unknowns to the integration level where the original domain is recovered. Due to the discontinuity introduced by the indicator function $\alpha(\boldsymbol{x})$, a standard Gauss-Legendre quadrature is not the optimal numerical integration scheme [59, 1]. A better integration scheme, also used in the results presented in section 4, is the space-tree approach where a recursive subdivision of cut cells is carried out to accurately integrate the discontinuous integrals. In 2D, the subdivision method of choice is the *quadtree*, in 3D the *octree*. The space-tree approach is very robust but generates a large number of quadrature points. Hence more sophisticated schemes have been developed for better efficiency: smart-octree [44, 45], moment-fitting [41], quadratic re-parameterization for tetrahedral FCM [73] and adaptively weighted quadratures [78].

### 3.3.2 Application of non-boundary conforming boundary conditions

In the FCM, the boundaries of the physical domain no longer conform with the finite element mesh. Therefore changes in the application of boundary conditions might be necessary as well. The application of non-homogeneous Neumann boundary conditions involves an evaluation of the corresponding boundary integral on the right hand side of the weak form. However, since the boundary is not resolved by the mesh in the FCM, Neumann boundary conditions must be applied by an explicit surface discretization. Dirichlet boundary conditions are imposed weakly, for example, using the penalty method or Nitsche's method [58]. These methods extend the weak formulation to include the Dirichlet boundary conditions in a weak sense [65]. In this work, the penalty approach is followed and its formulation is presented in section 3.3.3. Similarly to the application of Neumann boundary conditions, the constraining expressions are evaluated using a separate surface integration mesh which does not introduce any additional degrees of freedom.

### 3.3.3 Finite cell formulation for phase-field modeling of quasi-static brittle fracture

In this section, the modifications necessary in the weak form for the finite cell implementation are discussed. Only the quasi-static brittle fracture case is considered. The counterpart of the weak momentum equation in Eq:(20) for the quasi-static case with the FCM is

$$(\boldsymbol{\sigma}, \nabla \boldsymbol{w})_{\Omega_{\text{phy}}} + (\epsilon\,\boldsymbol{\sigma}, \nabla \boldsymbol{w})_{\Omega_{\text{fict}}} + (\beta \boldsymbol{u}, \boldsymbol{w})_{\partial \Omega_g}$$
$$= (\rho \boldsymbol{b}, \boldsymbol{w})_{\Omega_{\text{phy}}} + (\boldsymbol{h}, \boldsymbol{w})_{\partial \Omega_h} + (\beta \boldsymbol{g}, \boldsymbol{w})_{\partial \Omega_g}. \quad (27)$$

The penalty method is used to impose the Dirichlet boundary conditions with $\beta$ as the penalty parameter. The value of this parameter is such that it is large enough to yield accurate results, but small enough to avoid extreme ill-conditioning. Penalty parameters are typically three to five orders of magnitude higher than the stiffness of the material.

Similarly, the counterpart of the weak phase-field equation in Eq:(20) for the quasi-static case and the FCM is

$$([\frac{4l}{G_c}(1-\eta)\mathcal{H}+1]s, q)_{\Omega_{\text{phy}}} + (\epsilon\,[\frac{4l}{G_c}(1-\eta)\mathcal{H}+1]s, q)_{\Omega_{\text{fict}}}$$
$$+(4l^2\nabla s, \nabla q)_{\Omega_{\text{phy}}} + (\epsilon\,4l^2\nabla s, \nabla q)_{\Omega_{\text{fict}}} = (1, q)_{\Omega_{\text{phy}}}. \quad (28)$$

Due to the homogeneous Neumann boundary conditions for the phase-field throughout the boundary, enforcing the boundary conditions is trivial.

## 4 Numerical examples

In this section, the numerical performance of the phase-field fracture model combined with the uniform multi-level $hp$-refinement and the FCM is investigated. Illustrative examples for quasi-static and dynamic brittle fracture are presented. The pre-existing crack is defined using a history variable similar to [15]. In each example, the history variable associated to the phase-field is discretized in a stepwise-constant manner on a uniform background grid. Integrated Legendre polynomials are used as basis functions for the discretization and a $p+1$ integration rule is adopted for numerical integration. Results with two staggered iterations are presented.

### 4.1 Single-edge notched quasi-static tension test

The single-edge notched tension test example has become a typical benchmark example in the phase-field literature (see [16, 52, 15, 4]). For this reason, the same example with the same material and geometric parameters as in [4] is repeated here to test the proposed adaptive refinement scheme. Consider a two-dimensional square plate of side 1 mm with a pre-existing horizontal crack at mid-height as shown in Fig.5. The notched plate is subjected to a uniform vertical displacement on its top edge. The material and model parameters used are: $E$ = 210 GPa, $\nu$ = 0.3 ($\lambda$ = 121.15 kN/mm$^2$ and $\mu$ = 80.77 kN/mm$^2$), $G_c$ = 0.0027 kN/mm, $l$ = 0.004 mm, $\eta$ = $10^{-6}$. A displacement increment $\Delta u = 1 \times 10^{-5}$mm is used. The intent here is to test the proposed adaptive refinement scheme and study the convergence behavior of the phase-field solution under two scenarios.

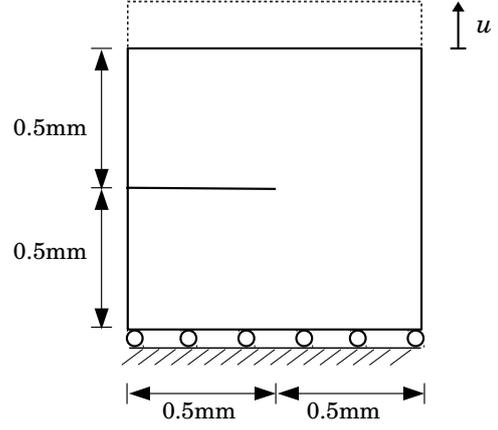

Figure 5: Geometry and boundary conditions for the single-edge notched tension test.

To this end, first, three discretizations with linear, quadratic and cubic ansatz functions and a grid of $4 \times 4$ base elements are chosen. Fig.6 depicts snapshots of the crack path at different displacements. It is to be noted how the refinement strategy keeps the refinement local around the mid-height of the plate where the crack is located and also how the discretization changes *automatically* as the crack advances. Qualitatively, one can observe in Fig.6 that the difference in the amount of crack propagation using quadratic and cubic ansatz is significantly less than that using linear and quadratic, indicating a converging behavior. This becomes clear in the corresponding load-displacement curve Fig.8(a), where a converging behavior is established as the order of the polynomial is increased. As expected, the peak force value using linear basis is higher and shifted towards a higher displacement value. We use the results presented by Ambati et al. [4] as the reference solution.

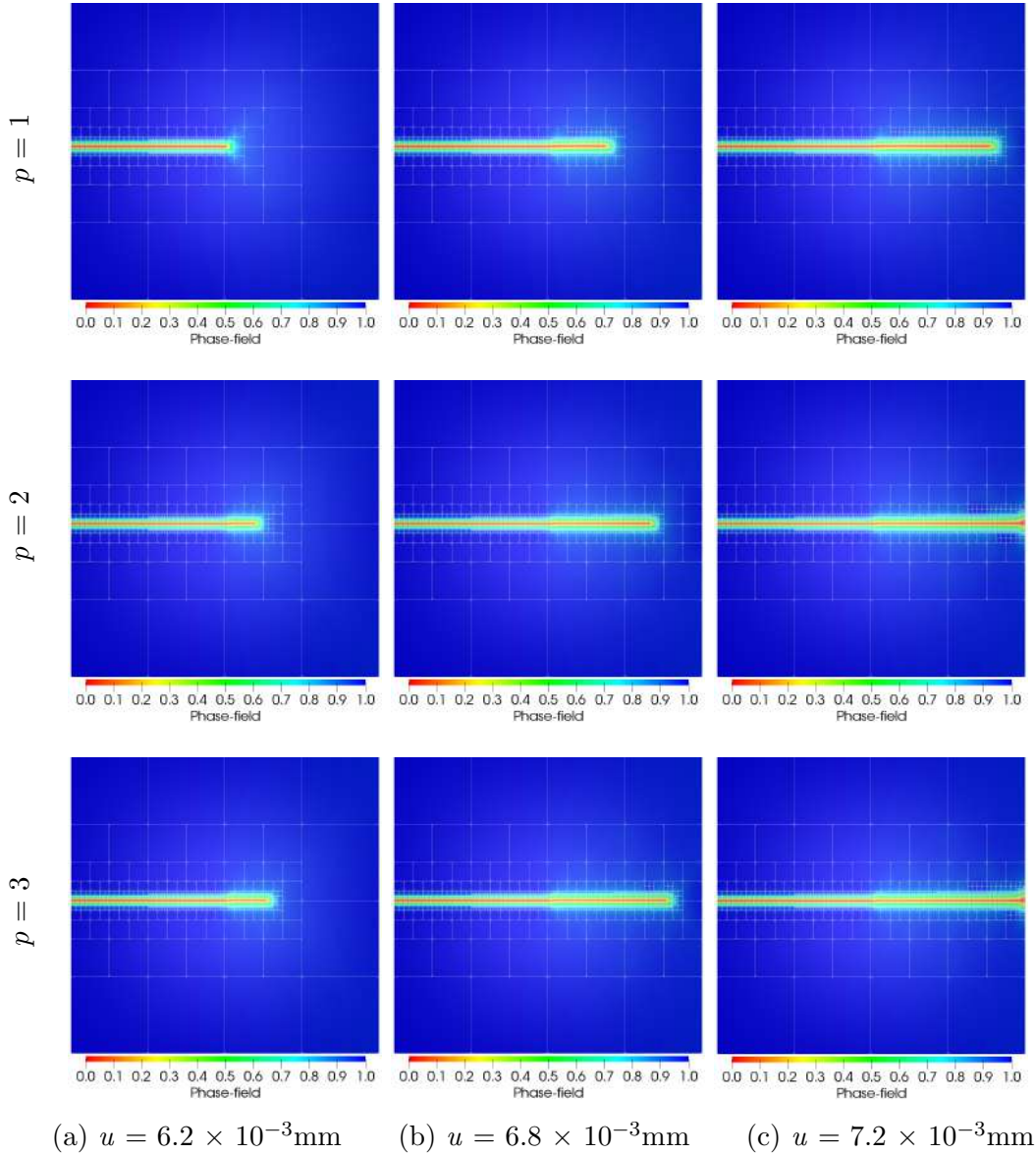

(a) $u = 6.2 \times 10^{-3}$mm     (b) $u = 6.8 \times 10^{-3}$mm     (c) $u = 7.2 \times 10^{-3}$mm

Figure 6: Single-edge notched tension test. Multi-level $hp$-refinement for different ansatz orders with $k=6$. Crack phase-field at different applied displacements.

To investigate the minimum refinement depth required for a given ansatz order, consider a grid with $4 \times 4$ elements, ansatz order $p = 3$ and three different depths $k = 4$, 5 and 6. Fig.7 depicts the phase-field evolution and Fig.8(b) shows the corresponding load-displacement curves. A significant difference is observed in the phase-field evolution for the lower refinement case, where the initial crack is not well defined as a consequence of the insufficient refinement depth, compared to the cases with higher refinement depths. For a refinement depth $k = 4$, the smallest element has a length that is almost four times the length-scale parameter due to which the crack propagation becomes rather slow with overestimated loads. This case thus emphasizes the importance of resolving the length-scale parameter. With increasing refinement depth, the length scale parameter is resolved better and hence the results rapidly converge to match the reference results.

It is clear from the above two studies that, with different ansatz orders and with different refinement depths, it is important to carefully handle the discretization in order to avoid mesh-related effects. For this example, a dynamically refined discretization with $4 \times 4$ elements, cubic shape functions and a refinement depth of 5 with around 7000 initial DOFs leading to around 18500 DOFs at complete fracture is sufficient to accurately capture the fracture process. To obtain similar quality results, 26058 DOFs have been used in the reference results [4].

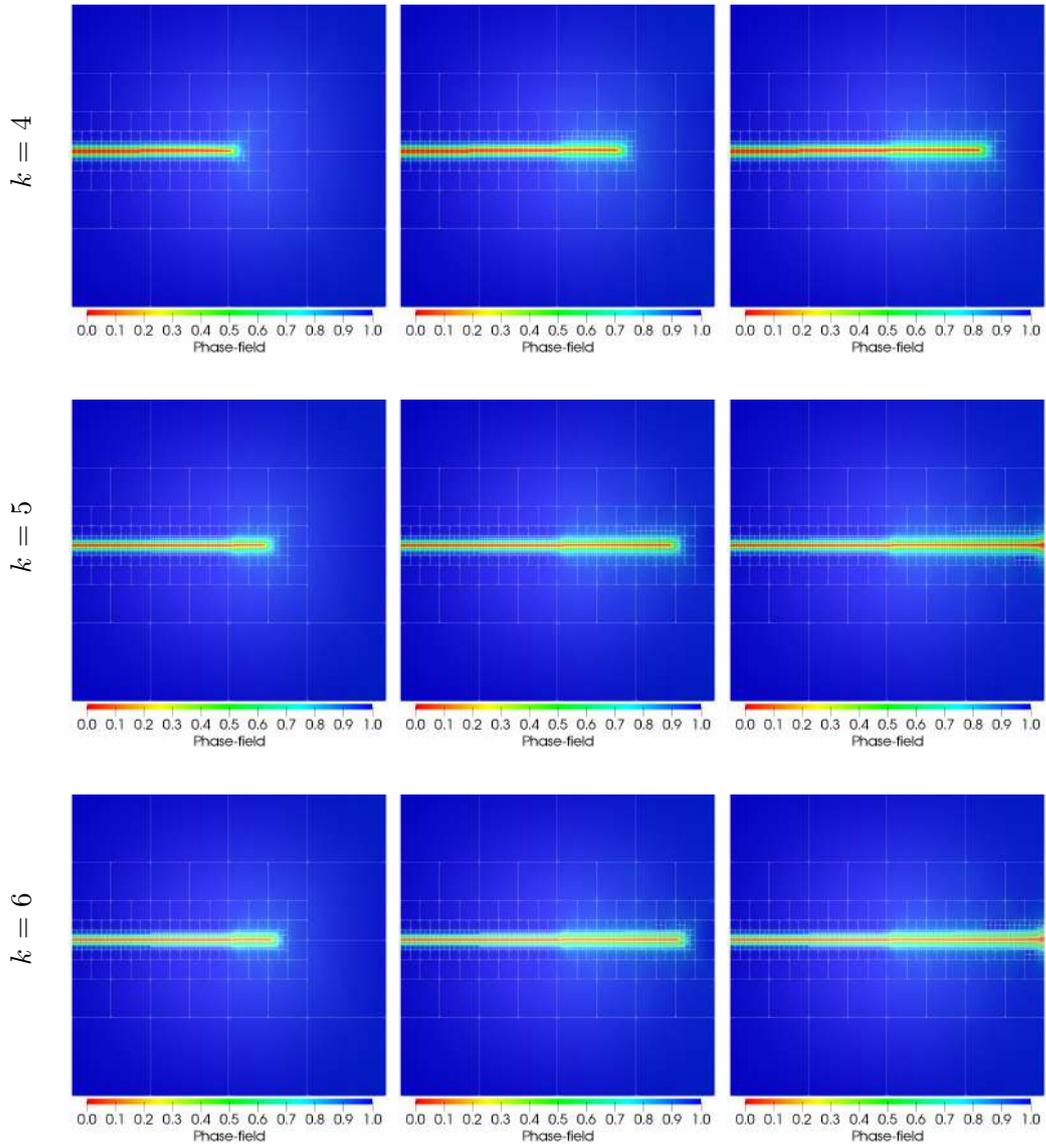

(a) $u = 6.2 \times 10^{-3}$mm  (b) $u = 6.8 \times 10^{-3}$mm  (c) $u = 7.0 \times 10^{-3}$mm

Figure 7: Single-edge notched tension test. Multi-level $hp$-refinement for different refinement depths with $p=3$. Crack phase-field evolution at different displacements.

## 4.2 Single-edge notched quasi-static shear test

With this example, a comparison between globally refined and locally refined discretizations for the same polynomial order of the shape functions is carried out. We consider the shear test specimen as shown in Fig.9, with the same material parameters as for the tension test specimen with a length scale parameter of $l = 0.01$ mm. A displacement increment $\Delta \boldsymbol{u} = 1 \times 10^{-4}$ mm is used.

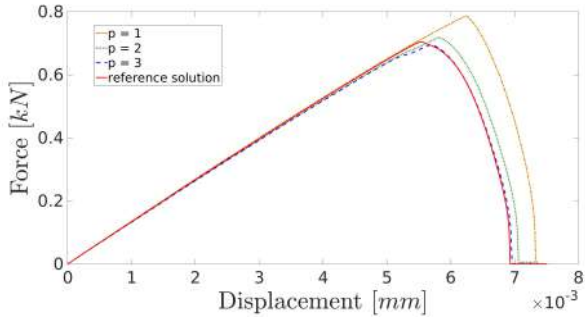 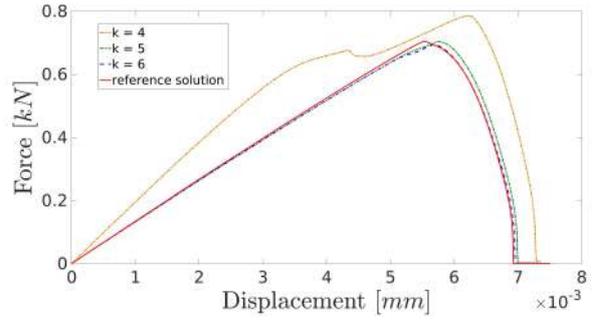

(a) Varying ansatz order $p$, fixed refinement depth $k=6$.   (b) Varying refinement depth $k$, fixed ansatz order $p=3$.

Figure 8: Single-edge notched tension test. Load-displacement curves with multi-level $hp$-refinement.

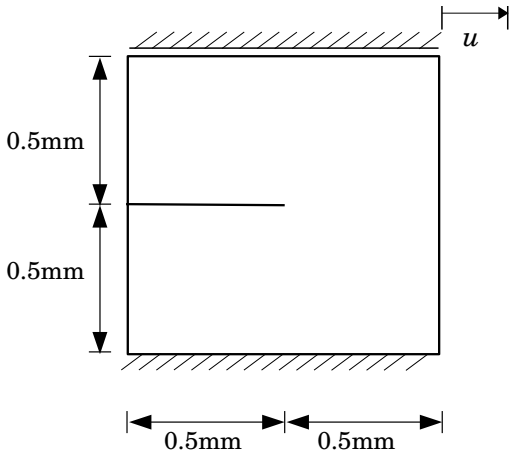

Figure 9: Geometry and boundary conditions for the single-edge notched shear test.

First, the specimen is uniformly discretized into a grid of 128 × 128 elements with linear ansatz, such that the smallest element is 0.0078 mm × 0.0078 mm in size. In order to compare the results obtained from the uniform mesh to those from a dynamically refined mesh using the uniform multi-level $hp$-refinement, we consider two cases: 32 × 32 elements with 2 multi-level $hp$-refinements and 16 × 16 elements with 3 multi-level $hp$-refinements. These two discretizations lead to the same dimensions for the smallest element as with the uniform grid. The structural response represented by the load-displacement curves is shown in Fig.10. Fig.11 shows the corresponding crack evolution for the three cases where the phase-field crack profile is virtually identical at the same displacement increments. The results clearly demonstrate that the same quality of the solution can be achieved in the vicinity of the crack by using a coarser base mesh if the refinement depth is increased appropriately. In Table 1, we report the number of DOFs and the run time[1] for the three cases under consideration. The results clearly indicate an increasing efficiency in terms of number of unknowns and run time as the refinement is made increasingly local. In other words, a smaller size of the system obtained using local refinements, that has a much lower computational time, allows for computations on larger domains.

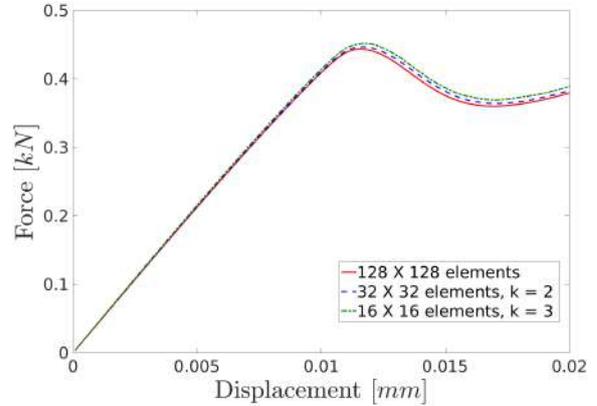

Figure 10: Single-edge notched shear test. Comparison of load-displacement curves for uniformly and adaptively refined meshes with $p=1$.

This example also demonstrates the robustness of the dynamic refinement criteria, which *automatically* refines the mesh towards the crack as it propagates. Unlike the previous example where the crack advances along an element edge, here we have crack propagation at a skew angle. Nevertheless, the simple refinement criteria performs robustly.

---

[1] The run time comparison is performed on a desktop workstation using an Intel®Core™ i7-3770K CPU with 32GB RAM

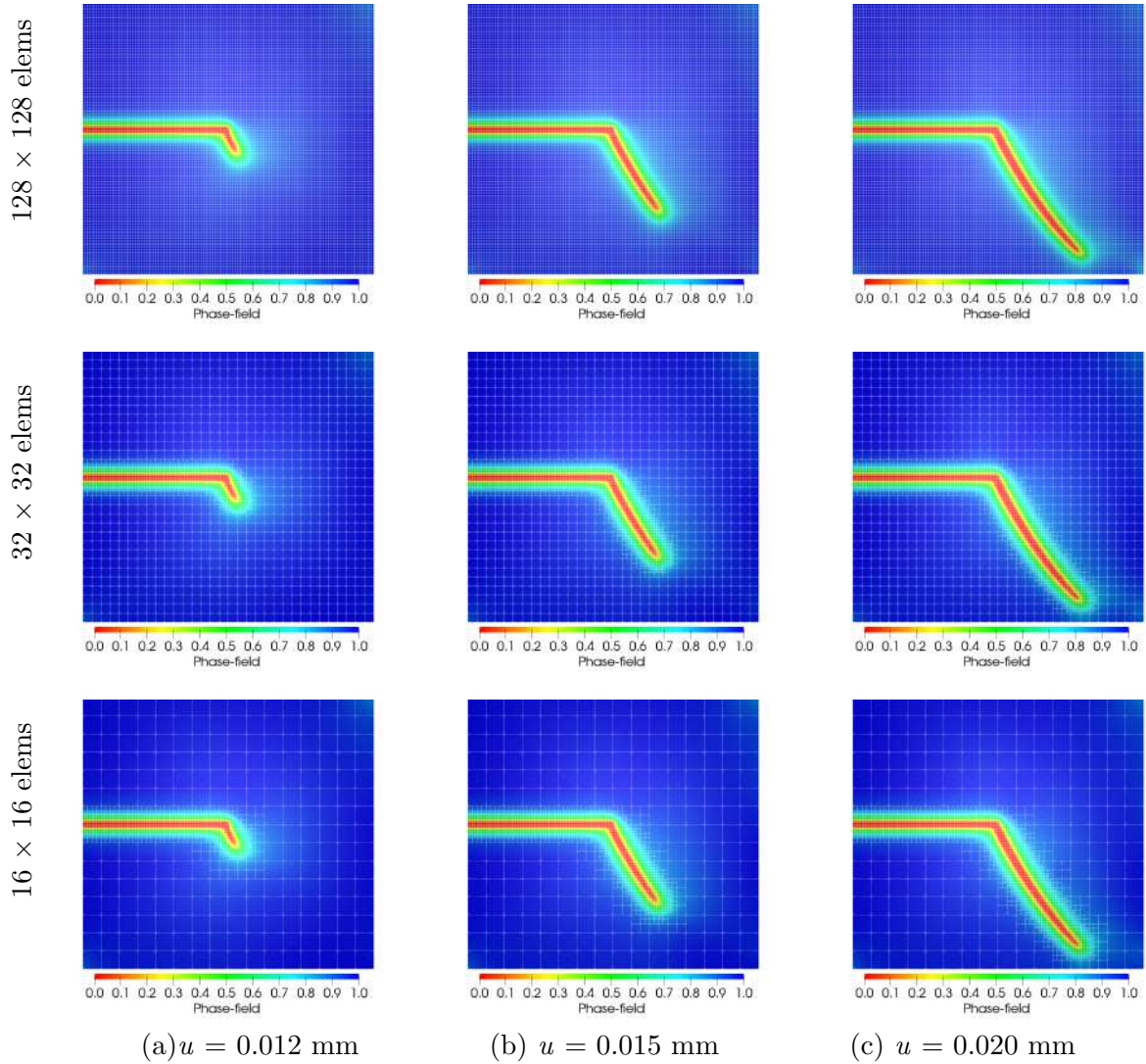

(a) $u = 0.012$ mm  (b) $u = 0.015$ mm  (c) $u = 0.020$ mm

Figure 11: Single-edge notched shear test. Comparison of crack phase-field for uniformly and adaptively refined meshes with $p=1$ at different applied displacements.

| Mesh [elements] | Multi-level $hp$-ref | Initial DOFs | Final DOFs | Run time [min] |
|---|---|---|---|---|
| 128×128 | 0 | 33828 | 33828 | 141 |
| 32×32 | 2 | 2678 | 5838 | 28 |
| 16×16 | 3 | 1168 | 4466 | 17 |

Table 1: Single-edge notched shear test. Parametric comparison for uniformly and adaptively refined meshes with $p=1$.

### 4.3 Dynamic crack branching under uniaxial tension

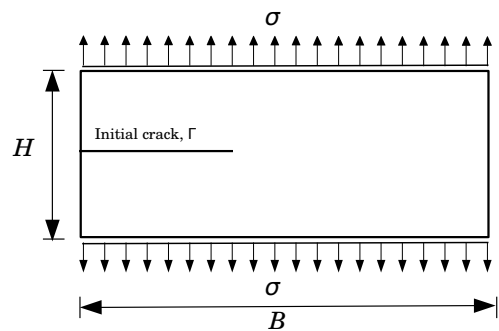

Figure 12: Geometry and loading of the dynamic uniaxial tension test.

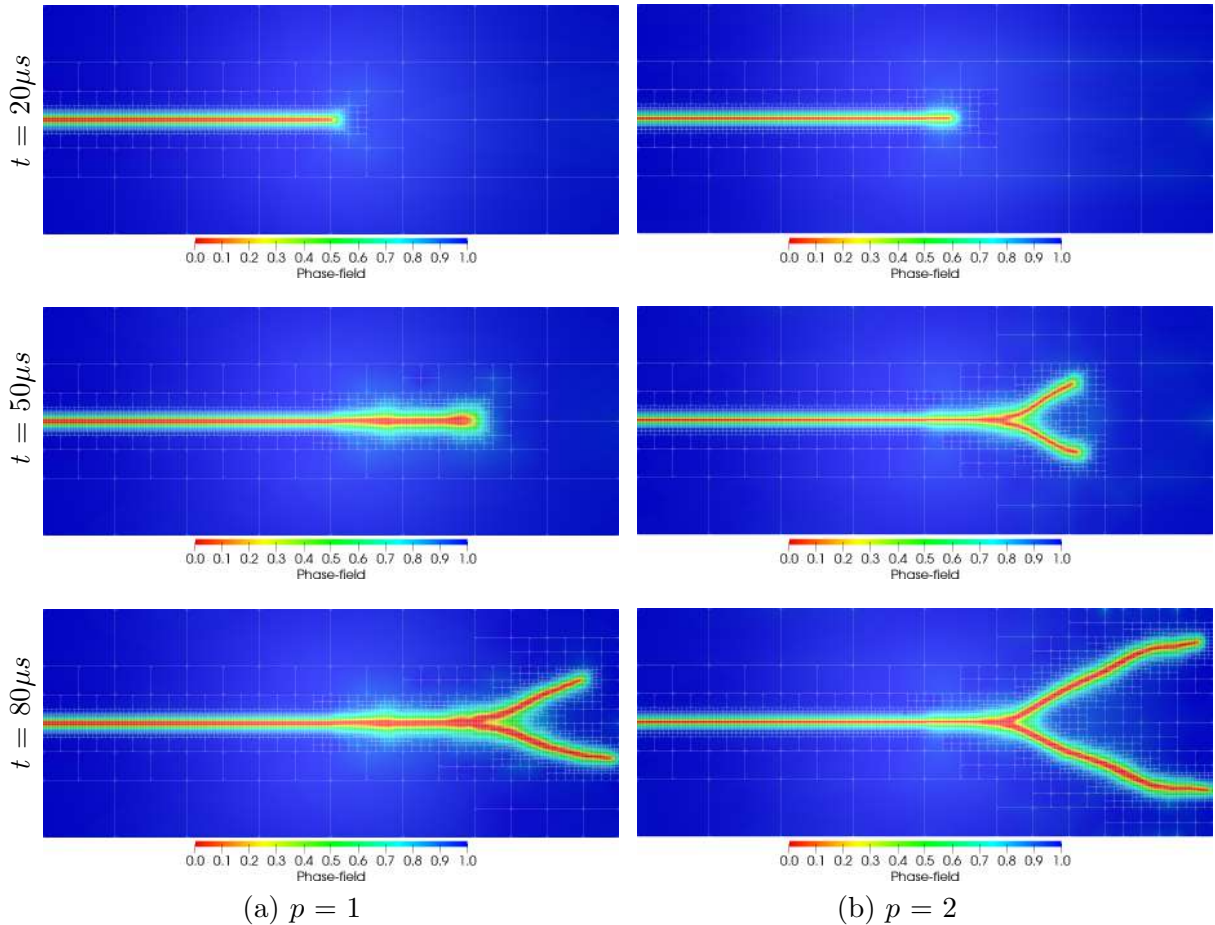

Figure 13: Dynamic crack branching under uniaxial tension. Multi-level $hp$-refinement for different ansatz orders with $k=5$. Crack phase-field evolution over time.

In this example (also studied in [75, 15]), a pre-notched rectangular specimen with length $B = 0.1$ m and $H = 0.04$ m is subjected to a tensile load $\sigma = 1$ MPa that remains constant in time as shown in Fig.12. This benchmark example truly is a difficult test case for the automatic refinement criterion as it involves the creation of crack branches over the course of the simulation. The material and model parameters used for simulation are $\rho = 2450$ kg/$m^3$, $E = 32$ GPa, $G_c = 3$ J/$m^2$, $\nu = 0.2$, $\eta = 10^{-6}$. A discretization with $8 \times 4$ elements with two different orders, linear and quadratic, time step $0.2$ $\mu$s and a length scale parameter of $l = 2.5 \times 10^{-4}$m are used.

When the problem is solved numerically, care should be taken while imposing the boundary conditions in order to avoid artifacts. If rigid body motion of the sample is not arrested, a singular or a nearly singular stiffness matrix is obtained, resulting in unphysical behavior. In order to restrict the rigid body motion, in the example presented here, the mid node on the right edge is fixed and a Dirichlet type constraint is weakly imposed on a small segment of the mid-line using the penalty method.

Crack evolution involving branching is depicted in Fig.13. The crack pattern obtained using the linear basis differs from that obtained using the quadratic basis. The latter results in the expected symmetrical crack pattern where the formation of branches is more pronounced and also occurs earlier. In Fig.14, the strain energy evolution over time obtained using the proposed adaptive refinement is compared that obtained by Borden et al [15] using a uniform mesh of size $h = 6.25 \times 10^{-5}$m. Similar to the observations made in the quasi-static case, convergence to the reference solution can be observed as the order of the shape functions is increased. Thus for a given number of staggered iterations, the quadratic basis produces more accurate results. However, a sudden jump in the strain energy can be observed at around 5μs when using the quadratic ansatz. This behavior is not completely understood and needs further investigation.

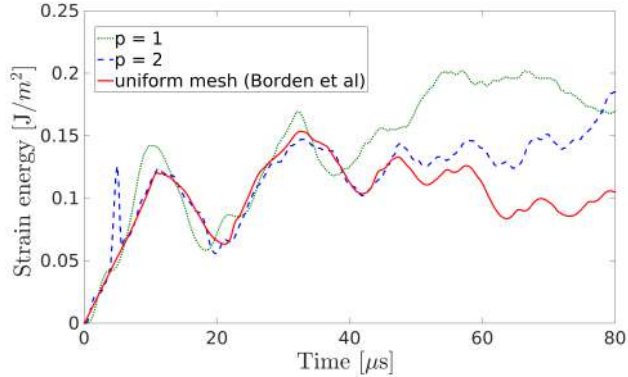

Figure 14: Dynamic crack branching under uniaxial tension. Multi-level $hp$-refinement for different ansatz orders with $k$=5. Plot of strain energy over time.

### 4.4 Notched plate with hole

This example demonstrates the combination of the FCM with phase-field modeling of brittle fracture. A notched plate with the geometry and boundary conditions as in Fig.15 is simulated. The material parameters used for this example are: $E = 6$ GPa, $\nu = 0.22$ ($\lambda = 1.94$ kN/mm$^2$ and $\mu = 2.45$ kN/mm$^2$), $G_c = 0.00228$ kN/mm, $l = 0.005$ mm, $\eta = 10^{-6}$ and a displacement increment $\Delta u = 5 \times 10^{-4}$mm is applied up to 0.1 mm. A discretization with $15 \times 30$ elements, quadratic shape functions and refinement depth $k = 4$ is used. The value of the indicator function is set to $10^{-8}$ in the fictitious domain, the three holes in this example. The quadtree integration scheme is used for the numerical integration with a partitioning depth of three.

As seen in Fig.15, a fixed displacement increment has to be imposed on the upper hole and a homogeneous displacement boundary condition on the lower hole. Here, since the notched plate with the three holes is embedded into a rectangle which is then discretized using a simple rectangular grid, we no longer have a geometry conforming mesh. This means, we no longer have conforming nodes on the perimeter of the upper and lower holes on which the boundary conditions can be applied. To resolve this, we create an explicit boundary corresponding to the holes and then impose the Dirichlet boundary conditions weakly on it using the penalty method as explained in sections 3.3.2 and 3.3.3.

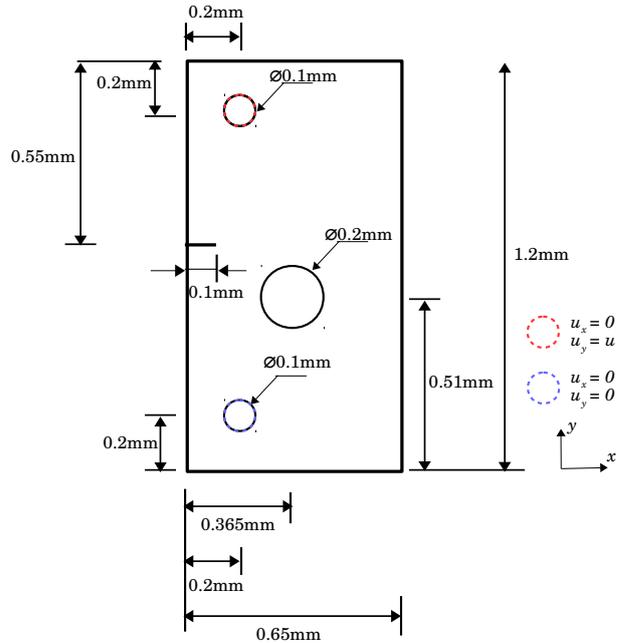

Figure 15: Geometry and boundary conditions of the notched plate with hole.

The obtained phase-field crack patterns at different displacement increments are shown in Fig.16. The results illustrate conspicuous benefits obtained by the novel combination of the FCM and the phase-field modeling of fracture. The crack propagates into the fictitious domain and re-emerges from the other side showing that the FCM modeling of the phase-field fracture evolution equation gives conceivable results, which are also comparable to the experimental results obtained for a similar specimen by Ambati et al. [4]. We highlight again the simple heuristic refinement that robustly follows the crack and surrounds it as soon as it re-emerges from the hole, as seen in the Fig.16(c).

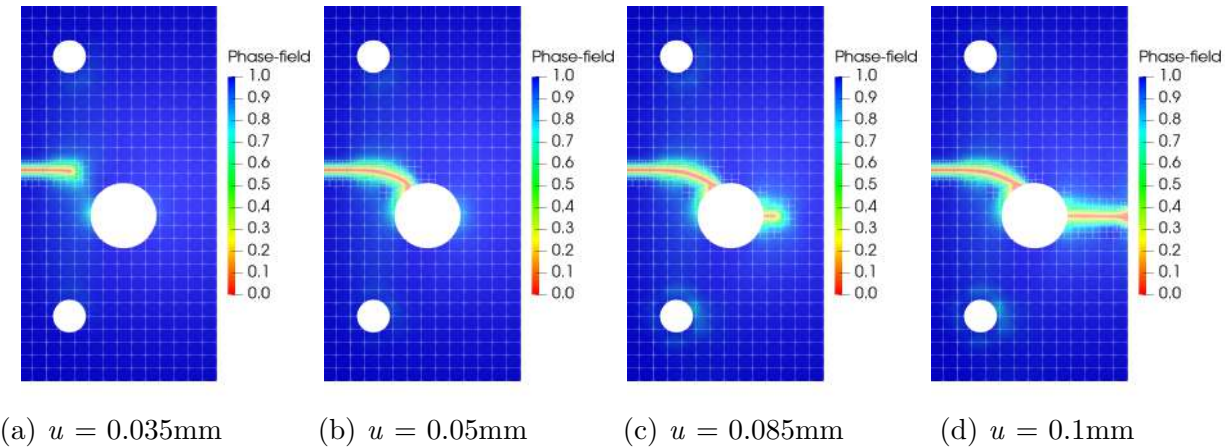

| (a) $u = 0.035$mm | (b) $u = 0.05$mm | (c) $u = 0.085$mm | (d) $u = 0.1$mm |

Figure 16: Notched plate with hole. Multi-level $hp$-refinement: $p$=2, $k$=4. Crack phase-field at different applied displacements.

## 5 Conclusions and outlook

The requirement of extremely fine meshes to resolve small length-scale parameters often limits the applicability of the phase-field approach to fracture problems. This issue can be solved by adaptively refining the mesh only where necessary. Additionally, high-order polynomials suit phase-field problems owing to the smoothness of the phase-field solution in most parts of the domain. These two ideas are combined in this work by integrating phase-field modeling of fracture with a multi-level $hp$-adaptive refinement scheme. The heuristic criterion developed to refine the mesh dynamically (over the course of the simulation) ensures $h$-refinement around the region with high gradients due to the presence of a crack. Since phase-field modeling eliminates the necessity of algorithmic tracking of the crack, the heuristic criterion remains numerically simple and robust. It can therefore be used in cases with complex crack profiles and easily be extended to three-dimensional problems. The benefits obtained from a local refinement scheme over a global refinement is clearly demonstrated by the numerical examples. The dynamic example gives evidence for the robustness of the refinement criterion. We also demonstrate the application of the finite cell method for phase-field modeling of fracture. The major advantages of using the finite cell approach instead of a classical finite element computation thus carry over to fracture problems: Highly complex geometric domains can readily be simulated without the need of generating a boundary conforming mesh. This benefit is expected to be particularly important in case of three-dimensional porous structures, which are a focus of our ongoing research.